\documentclass[a4paper]{article}

\usepackage{icrc2013}

\newcommand{\apj}{ApJ}

\newcommand{\aap}{A{\&}A}

\newcommand{\prd}{PhysicalReviewD}
\newcommand{\mnras}{MNRAS}

\newcommand{\hess}{H.E.S.S.}

\newcommand{\chandra}{\emph{Chandra}}

\newcommand{\fermi}{\emph{Fermi}}

\newcommand{\terzan}{Terzan\,5}
\newcommand{\htfive}{HESS\,J1747$-$248}

\title{Search for Very-high-energy $\gamma$-ray emission from Galactic globular clusters with \hess}

\shorttitle{Search for $\gamma$-ray emission from Galactic GCs with \hess}

\authors{
P.~Eger$^{1,2}$,
C.~van Eldik$^{2}$
for the \hess\ Collaboration.
}

\afiliations{
$^1$ Max-Planck-Institut f\"ur Kernphysik, Heidelberg, Germany \\
$^2$ Universit\"at Erlangen-N\"urnberg, Physikalisches Institut, Erlangen, Germany \\
}

\email{peter.eger@mpi-hd.mpg.de}

\abstract{
Globular clusters (GCs) are established emitters of high-energy (HE, 100 MeV$<$E$<$100 GeV) $\gamma$-ray radiation which could originate from the cumulative emission of the numerous millisecond pulsars (msPSRs) in the clusters' cores or from inverse Compton (IC) scattering of relativistic leptons accelerated in the GC environment. 
GCs could also constitute a new class of sources in the very-high-energy (VHE, E$>$100 GeV) $\gamma$-ray regime, judging from the recent detection of emission from the direction of Terzan 5 with the H.E.S.S. telescope array. 

To search for VHE $\gamma$-ray sources associated with other GCs, and to put constraints on leptonic emission models, we systematically analyzed the observations towards 15 GCs taken with H.E.S.S. 
We searched for individual sources of VHE $\gamma$-rays from each GC in our sample and also performed a stacking analysis combining the data from all GCs to investigate the hypothesis of a population of faint emitters. 
Assuming IC emission as the source of emission from Terzan 5, we calculated the expected $\gamma$-ray flux for each of the 15 GCs, based on their number of millisecond pulsars, their optical brightness and the energy density of background photon fields. 

We did not detect significant emission from any of the 15 GCs. 
The obtained flux upper limits allow to rule out the simple IC/msPSR scaling model for NGC6388 and NGC7078. 
The upper limits derived from the stacking analyses are factors between 2 and 50 below the flux predicted by the simple leptonic model, depending on the assumed source extent and the dominant target photon fields. 
Therefore, Terzan 5 still remains exceptional among all GCs, as the VHE $\gamma$-ray emission either arises from extra-ordinarily efficient leptonic processes, or from a recent catastrophic event, or is even unrelated to the GC itself.
}


\begin{document}
\maketitle

\section{Introduction}
\label{sec-introduction}
Globular clusters (GCs) are very dense systems of stars, hosting the oldest and most evolved stellar populations of the Galaxy. 
The extremely high rate of close stellar encounters in the cores of GCs likely gives rise to large numbers of dynamically formed compact binary systems (see e.g.\cite{2006ApJ...646L.143P}) which are believed to be the  progenitors of millisecond pulsars (msPSRs). 
Indeed, large msPSR populations have been detected in many Galactic GCs through radio observations  \cite{2005ASPC..328..147C}. 
Because msPSRs can be strong sources of pulsed magnetospheric emission in the high-energy (HE, 100\,MeV$<E<$100\,GeV) $\gamma$-ray regime (as shown by observations with \fermi /LAT  \cite{2009Sci...325..848A}), $\gamma$-ray signals are expected from GCs either due to bright individual msPSRs or due to the cumulative emission from the whole population.

A complementary approach to access even the whole msPSR population in GCs could come through observations of very-high-energy (VHE, $E>$100\,GeV) $\gamma$-rays with ground-based imaging atmospheric Cherenkov telescopes, such as the high energy stereoscopic system (\hess ). 
Models predicting emission in the VHE regime rely on inverse Compton (IC) up-scattering of soft photon fields by ultra-relativistic leptons accelerated in the pulsars' magnetospheres \cite{2009ApJ...696L..52V} or even re-accelerated in colliding pulsar wind nebula shocks \cite{2007MNRAS.377..920B}. 
The relevant target photon fields for the IC process depend on the diffusion properties of the relativistic lepton population. 
If the particles are contained in the cores of the GCs the starlight of the member stars would be dominant. 
On the other hand, for scenarios involving fast diffusion away from the acceleration site other external photon fields such as the cosmic microwave background or Galactic background fields would become important. 

Recently, VHE $\gamma$-ray emission has been detected from the direction of the GC \terzan\ with \hess\ (\htfive , \cite{2011A&A...531L..18H}). 
Should this signal indeed be associated to \terzan\ this would establish GCs as a new class of sources in VHE $\gamma$-rays. 
In the case of IC emission also diffuse X-ray source could be expected, arising from synchrotron radiation (SR) from the same population of leptons \cite{2008AIPC.1085..277V}. 
Interestingly, extended and diffuse non-thermal X-ray emission was detected from \terzan\ with \chandra\ by \cite{2010A&A...513A..66E}.
As an alternative to the leptonic msPSR scenarios, a hadronic model was proposed for \htfive\ by \cite{2011A&A...533L...5D}. 
Here, it is suggested that the VHE $\gamma$-ray emission arises from the decay of neutral pions that were produced in inelastic collisions of relativistic protons with interstellar matter.
These protons could have been accelerated in a remnant of a short $\gamma$-ray burst resulting from a merger of binary neutron stars, which are very common in GC cores. 

In this work we present a systematic search for VHE $\gamma$-ray emission from all Galactic GCs that are covered by the present observational data set of the \hess\ telescope array. 
To increase the sensitivity for the detection of a population of individually faint emitters we also performed a stacking analysis, where we combined the number of photon candidate counts from the individual GCs to test for a statistically significant signal.
We then compare the obtained results to the VHE $\gamma$-ray signal detected from \terzan\ to investigate scenarios based on IC emission from leptons accelerated by msPSRs. 
For more details on the analyses and interpretations presented in this study, please refer to \cite{2013A&A...551A..26H}. 

\section{\hess\ observations}
\hess\ is an array of four imaging atmospheric Cherenkov telescopes located in the Khomas Highland of Namibia. 
Each telescope is equipped with a tessellated mirror with a surface area of 107\,m$^2$ and a camera comprising 960 photomultiplier tubes with a total field of view (FoV) of 5$^\circ$. 
With modern reconstruction techniques the \hess\ array has a point-source sensitivity of about 2$\times 10^{-13}$\,ph\,cm$^{-2}$\,s$^{-1}$ (E$>$200\,GeV) for 25 hours of observation, requiring a statistical significance of 5\,$\sigma$ \cite{2009APh....32..231D,2009APh....31..383O}. 

\begin{table}[tb]
\caption[]{The GC sample studied in this work}
\renewcommand{\tabcolsep}{2.5pt}
\begin{center}
\begin{tabular}{lllll}
\hline\hline\noalign{\smallskip}
\multicolumn{1}{l}{GC} &
\multicolumn{1}{l}{long.$^{(1)}$} &
\multicolumn{1}{l}{lat.$^{(1)}$} &
\multicolumn{1}{l}{zenith$^{(2)}$} &
\multicolumn{1}{l}{livetime$^{(3)}$} \\
\multicolumn{1}{l}{name} &
\multicolumn{1}{l}{($^{\circ}$)} &
\multicolumn{1}{l}{($^{\circ}$)} &
\multicolumn{1}{l}{($^{\circ}$)} &
\multicolumn{1}{l}{(h)} \\
\noalign{\smallskip}\hline\noalign{\smallskip}
NGC\,104 (47\,Tuc)$^{(*)}$        &       305.89  &       $-$44.89        &       49.8    &              23.1    \\
NGC\,6388$^{(*)}$       &       345.56  &       $-$6.74 &       23.6        &       17.9    \\
NGC\,7078 (M\,15)       &       65.01   &       $-$27.31        &       37.5    &      12.3    \\
Terzan\,6 (HP\,5)       &       358.57  &       $-$2.16 &       24.6       &       15.2    \\
Terzan\,10&     4.49    &       $-$1.99 &       18.1     &       4.2             \\
NGC\,6715 (M\,54)      &       5.61    &       $-$14.09        &       18.3    &         11.8    \\
NGC\,362        &       301.53  &       $-$46.25        &       49.5       &       5.0             \\
Pal\,6  &       2.10    &       1.78    &       19.4    &       24.7    \\
NGC\,6256       &       347.79  &       3.31    &       20.5    &        5.3             \\
Djorg\,2        &       2.77    &       $-$2.50 &       15.9    &           4.6     \\
NGC\,6749       &       36.20   &       $-$2.21 &       33.7    &           8.2     \\
NGC\,6144       &       351.93  &       15.70   &       26.8    &           4.7             \\
NGC\,288        &       152.30  &       $-$89.38        &       11.8    &          46.7    \\
HP\,1 (BH\,229)          &       357.44  &       2.12    &       14.3    &          5.6     \\
Terzan\,9       &       3.61    &       $-$1.99 &       18.4       &       5.2     \\
\hline\noalign{\smallskip}
\end{tabular}
\label{tab-gc-sample}
\end{center}
$^{(1)}$Galactic coordinates used for the analyses, as given by H10;
$^{(2)}$Mean zenith angle and offset averaged over all contributing runs;
$^{(3)}$Acceptance corrected livetime of all \hess\ observation runs passing quality cuts;
$^{(*)}$GC is detected with \fermi /LAT.
\end{table}

We based our target selection on the GC catalog of \cite{1996AJ....112.1487H} from which we also took all basic parameters such as position, distance ($d$), core radius ($r_\mathrm{c}$), and central luminosity density ($\rho_\mathrm{0}$). 
To select the sample of GCs and observational data that are suited for our study with \hess\ we required a 
minimum distance of the GC to the Galactic plane of 1.0$^\circ$ to avoid chance coincidences with unrelated sources. 
Also, we required a minimum exposure of 20 good quality runs ($\sim$28 minutes each).
Table~\ref{tab-gc-sample} summarizes the sample of the 15 GCs passing our \emph{a priori} cuts. 

\begin{table*}[t]
\caption[]{Analysis results and model predictions}
\begin{center}
\begin{tabular}{llllllllll}
\hline\hline\noalign{\smallskip}
\multicolumn{1}{l}{GC} &
\multicolumn{1}{l}{E$_\mathrm{th}^{(1)}$} &
\multicolumn{1}{l}{${N_\mathrm{ON}}^{(2)}$} &
\multicolumn{1}{l}{${N_\mathrm{OFF}}^{(2)}$} &
\multicolumn{1}{l}{1/$\alpha^{(3)}$} &
\multicolumn{1}{l}{sig.$^{(4)}$} &
\multicolumn{1}{l}{$r^{(5)}$} &
\multicolumn{1}{l}{${F_\mathrm{UL}(E>\mathrm{E}_\mathrm{th})}^{(6)}$} &
\multicolumn{1}{l}{${F_\mathrm{UL} / F_\mathrm{IC;GC}}^{(7)}$} &
\multicolumn{1}{l}{${F_\mathrm{UL} / F_\mathrm{IC;IR,opt,CMB}}^{(7)}$} \\
\multicolumn{1}{l}{name} &
\multicolumn{1}{l}{(TeV)} &
\multicolumn{2}{c}{(counts)} &
\multicolumn{1}{l}{} &
\multicolumn{1}{l}{($\sigma$)} &
\multicolumn{1}{l}{($^\circ$)} &
\multicolumn{1}{c}{(ph\,cm$^{-2}$\,s$^{-1}$)} &
\multicolumn{2}{l}{} \\
\noalign{\smallskip}\hline\noalign{\smallskip}
\multicolumn{4}{l}{\emph{a) point-like source analysis}} \\
NGC\,104        &       0.72    &       72      &       941     &       18.2    &       2.6             &--& 1.9$\times 10^{-12}$    &       2.6$\times 10^{-1}$       &       2.1$\times 10^{1}$    \\
NGC\,6388       &       0.28    &       180     &       2365    &       14.9    &       1.6             &--& 1.5$\times 10^{-12}$    &       8.0$\times 10^{-2}$       &       1.6$\times 10^{0}$    \\
NGC\,7078       &       0.40    &       119     &       1988    &       15.0    &       $-$1.2          &--& 7.2$\times 10^{-13}$    &       1.9$\times 10^{-1}$       &       2.1$\times 10^{1}$    \\
Terzan\,6       &       0.28    &       202     &       8194    &       42.0    &       0.5             &--& 2.1$\times 10^{-12}$    &       7.3$\times 10^{-1}$       &       1.0$\times 10^{0}$    \\
Terzan\,10      &       0.23    &       76      &       2455    &       36.0    &       0.9             &--& 2.9$\times 10^{-12}$    &       4.3$\times 10^{-1}$       &       2.7$\times 10^{-1}$   \\
NGC\,6715       &       0.19    &       159     &       2361    &       15.2    &       0.3             &--& 9.3$\times 10^{-13}$    &       3.1$\times 10^{-1}$       &       1.3$\times 10^{2}$    \\
NGC\,362        &       0.59    &       18      &       533     &       33.0    &       0.4             &--& 2.4$\times 10^{-12}$    &       3.9$\times 10^{0}$        &       1.8$\times 10^{2}$   \\
Pal\,6          &       0.23    &       363     &       10810   &       31.4    &       1.0             &--& 1.2$\times 10^{-12}$    &       1.3$\times 10^{1}$        &       1.1$\times 10^{1}$   \\
NGC\,6256       &       0.23    &       64      &       1869    &       27.4    &       $-$0.5          &--& 3.2$\times 10^{-12}$    &       1.8$\times 10^{1}$        &       2.9$\times 10^{1}$   \\
Djorg\,2        &       0.28    &       56      &       2387    &       39.4    &       $-$0.6          &--& 8.4$\times 10^{-13}$    &       1.0$\times 10^{1}$        &       1.0$\times 10^{1}$   \\
NGC\,6749       &       0.19    &       84      &       2633    &       29.3    &       $-$0.6          &--& 1.4$\times 10^{-12}$    &       2.5$\times 10^{1}$        &       4.1$\times 10^{1}$   \\
NGC\,6144       &       0.23    &       63      &       2196    &       30.8    &       $-$1.0          &--& 1.4$\times 10^{-12}$    &       3.8$\times 10^{2}$        &       1.1$\times 10^{3}$   \\
NGC\,288        &       0.16    &       647     &       24148   &       38.5    &       0.8             &--& 5.3$\times 10^{-13}$    &       2.7$\times 10^{2}$        &       3.2$\times 10^{3}$   \\
HP\,1           &       0.23    &       67      &       2771    &       34.3    &       $-$1.6          &--& 1.5$\times 10^{-12}$    &       5.2$\times 10^{2}$        &       1.7$\times 10^{2}$   \\
Terzan\,9       &       0.33    &       89      &       2556    &       31.7    &       0.9             &--& 4.5$\times 10^{-12}$    &       2.6$\times 10^{4}$        &       9.0$\times 10^{2}$   \\
\hline\noalign{\smallskip}
\multicolumn{4}{l}{\emph{b) extended source analysis}}\\
NGC\,104        &       "        &   293 &   2016        &   7.4         &   1.2        & 0.22	 &	2.3$\times 10^{-12}$	 &	 2.3$\times 10^{-1}$	  &	  1.9$\times 10^{1}$	  \\
NGC\,6388       &       "        &   253 &   2818        &   12.9        &   2.2        & 0.11	 &	1.7$\times 10^{-12}$	 &	 9.2$\times 10^{-2}$	  &	  1.8$\times 10^{0}$	  \\
NGC\,7078       &       "        &   161 &   2386        &   14.0        &   $-$0.7     & 0.11	 &	1.1$\times 10^{-12}$	 &	 2.8$\times 10^{-1}$	  &	  3.1$\times 10^{1}$	  \\
Terzan\,6       &       "        &   304 &   9802        &   34.2        &   1.0        & 0.12	 &	2.4$\times 10^{-12}$	 &	 8.1$\times 10^{-1}$	  &	  1.2$\times 10^{0}$	  \\
Terzan\,10      &       "        &   218 &   4134        &   19.0        &   0.0        & 0.18	 &	3.6$\times 10^{-12}$	 &	 5.4$\times 10^{-1}$	  &	  3.4$\times 10^{-1}$	    \\
NGC\,6715       &       "    	 &   159 &   2361        &   15.2        &   0.3        & *      &      9.3$\times 10^{-13}$    &       3.1$\times 10^{-1}$       &       1.3$\times 10^{2}$    \\
NGC\,362        &       "        &   30  &   708         &   25.6        &   0.4        & 0.13	 &	2.5$\times 10^{-12}$	 &	 4.0$\times 10^{0}$	 &	 1.8$\times 10^{2}$	 \\
Pal\,6          &       "        &   1148&   17631       &   16.6        &   2.5        & 0.18 	 &	2.1$\times 10^{-12}$	 &	 2.4$\times 10^{1}$	 &	 1.9$\times 10^{1}$	 \\
NGC\,6256       &       "        &   131 &   2524        &   20.4        &   0.6        & 0.13	 &	3.9$\times 10^{-12}$	 &	 2.1$\times 10^{1}$	 &	 3.5$\times 10^{1}$	 \\
Djorg\,2        &       "        &   137 &   3753        &   24.8        &   $-$1.2     & 0.16	 &	9.7$\times 10^{-13}$	 &	 1.2$\times 10^{1}$	 &	 1.2$\times 10^{1}$	 \\
NGC\,6749       &       "        &   168 &   3544        &   20.7        &   $-$0.3     & 0.14	 &	2.1$\times 10^{-12}$	 &	 3.6$\times 10^{1}$	 &	 5.9$\times 10^{1}$	 \\
NGC\,6144       &       "        &   120 &   2913        &   23.9        &   $-$0.2     & 0.13	 &	2.5$\times 10^{-12}$	 &	 6.7$\times 10^{2}$	 &	 1.9$\times 10^{3}$	 \\
NGC\,288        &       "        &   1030&   30767       &   30.7        &   0.8        & 0.13	 &	6.1$\times 10^{-13}$	 &	 3.1$\times 10^{2}$	 &	 3.7$\times 10^{3}$	 \\
HP\,1           &       "        &   67  &   2771        &   34.3        &   $-$1.6     & *   	 &	1.5$\times 10^{-12}$	 &	 5.2$\times 10^{2}$	 &	 1.7$\times 10^{2}$	 \\
Terzan\,9       &       "        &   206 &   3909        &   18.8        &   $-$0.1     & 0.16	 &	4.1$\times 10^{-12}$	 &	 1.8$\times 10^{4}$	 &	 6.2$\times 10^{2}$	 \\
\hline\noalign{\smallskip}
\multicolumn{9}{l}{\emph{stacking analysis}}\\
\emph{a)}       &       0.23    &       2242    &       67826   &       31.2    &       1.6     &       --      &       3.3$\times 10^{-13}$    &       (5.4$^{+16}_{-1.7}$)$\times 10^{-2}$     &       (4.3$^{+11}_{-1.4}$)$\times 10^{-1}$     \\[3pt]
\emph{b)}       &       "       &       4425    &       92037   &       21.6    &       2.4     &       --      &       4.5$\times 10^{-13}$    &       (7.5$^{+23}_{-2.4}$)$\times 10^{-2}$     &       (5.9$^{+17}_{-2.0}$)$\times 10^{-1}$     \\[3pt]
\hline\noalign{\smallskip}
\end{tabular}
\label{tab-analysis-results}
\end{center}
$^{(1)}$Energy threshold of the analysis, defined as the location of the peak in the distribution of reconstructed photon energies;
$^{(2)}$Total number of on- and off-counts; 
$^{(3)}$Ratio between off- and on-exposure when applying the \emph{reflected background} technique;
$^{(4)}$Detection significance (pre-trial) following \cite{1983ApJ...272..317L};
$^{(5)}$Extraction radius used for the analysis assuming extended emission; a star(*) denotes the cases where the calculated intrinsic extent is negligible compared to the PSF;
$^{(6)}$Photon flux upper limits (99\% c.l., following \cite{1998PhRvD..57.3873F} assuming a power-law spectrum with an index of $-$2.5;
$^{(7)}$Ratio between the upper limit and the expected IC flux using either the GC starlight (GC), or the sum of the Galactic infrared (IR), Galactic optical (opt) and the CMB background as target photon fields. 
\end{table*}

\subsection{Analysis of individual GCs}
For the analysis we used the \emph{model} technique where the air showers are described by a semi-analytical model. 
The expected camera images are then compared to the observational data based on a maximum likelihood method \cite{2009APh....32..231D}. 
The \emph{model} analysis yields an improved sensitivity, particularly at lower energies, and a more efficient Gamma-Hadron separation compared to the standard Hillas approach. 
We used standard cuts for the \emph{model} analysis which include a 60 photo electron cut on the image size. 
With this configuration the \emph{model} analysis features a PSF with a 68\% containment radius of 0.07$^\circ$ on average. 
We cross-checked all results with an independent calibration using a Hillas analysis framework, which employs a machine-learning algorithm based on Boosted Decision Trees \cite{2009APh....31..383O}. 

To search for VHE $\gamma$-ray emission from each GC we performed two different kinds of analyses where the size of the emission region was assumed to be either a) point-like (0.1$^\circ$ integration radius) or b) extended while using the nominal GC position as the center of the extraction region (see Tab.~\ref{tab-gc-sample}). 
For case b) we determined the expected intrinsic size (with the effects of the PSF removed) of the source related to each GC based on the measured extent of the VHE $\gamma$-ray source \htfive\ detected from the direction of \terzan\ \cite{2011A&A...531L..18H}. 
Assuming that the physical size of the VHE $\gamma$-ray source is the same among all GCs, we scaled the angular size of the major axis of \htfive\ with the relative distance of each individual GC (as given by H10) and \terzan\ ($d_\mathrm{T5} = 5.9$\,kpc, \cite{2009Natur.462..483F}). 
For the analyses we used the \emph{reflected background} technique \cite{2007A&A...466.1219B}, which ensures that the regions used for signal (ON) and background (OFF) extraction have the same acceptance in the FoV of the camera. 
In Tab.~\ref{tab-analysis-results} we present the results for both analyses and for each individual GC. 
From none of the GCs, neither with analysis a) nor b), significant excess emission is seen above the estimated background, as the individual significances for a detection, even pre-trials, are well below the threshold of 5\,$\sigma$. 
We derived upper limits for the photon flux above the energy threshold (99\% confidence level, following \cite{1998PhRvD..57.3873F}) assuming a power-law spectrum for the photon flux with an index of $\Gamma = -2.5$, which we chose to allow for a comparison with the results obtained for \terzan\ ($\Gamma_\mathrm{T5} = -2.5 \pm 0.3_\mathrm{stat} \pm 0.2_\mathrm{sys}$, \cite{2011A&A...531L..18H}).

\subsection{Stacking analysis}
Because no individual GC studied in this sample shows significant VHE $\gamma$-ray emission, we performed a stacking analysis to search for a population of faint emitters. 
To combine the analysis results and count-statistics from the observation runs of all GCs we applied the same method that is used to combine the results for analyses of individual sources. 
The total GC stack has an acceptance-corrected livetime of 195 hours of good quality data and features an energy threshold of 0.23\,TeV. 
We performed the stacking analysis for both the point-like and extended source analyses described in the previous section. 
The results are shown in Tab.~\ref{tab-analysis-results}. 
The total detection significances obtained with these stacking analyses are 1.6\,$\sigma$ and 2.4\,$\sigma$, respectively, which are also very well below the threshold of 5\,$\sigma$ required for a firm detection. 
Therefore, we calculated upper limits on the photon flux for the full stack, again assuming a power-law spectrum with an index of $\Gamma = -2.5$ (see Tab.~\ref{tab-analysis-results}).

\section{Discussion}
Even though neither any individual GC nor the stack as a whole revealed a significant $\gamma$-ray signal, the resulting upper limits on the photon flux can still be used to investigate models based on IC emission from leptons accelerated by msPSRs. 
Assuming that \htfive\ is indeed related to \terzan\ and that the msPSR/IC scenario is the dominant process for VHE $\gamma$-ray emission, the IC flux should scale as: 
\begin{equation}
\label{eq-IC-flux}
        F_\mathrm{IC} \propto N_\mathrm{msPSR} \times N_\mathrm{ph} \times d^{-2},
\end{equation}
with the number of msPSRs $N_\mathrm{msPSR}$, the total number of available target photons in the volume occupied by relativistic leptons for the IC process $N_\mathrm{ph}$ and the distance to the GC $d$ (Eq.~\ref{eq-IC-flux} is essentially the same as Eq.~8 in \cite{2008AIPC.1085..277V}). 
This simple scaling relation is of course only valid if the mechanisms for lepton production are similar among all GCs and, particularly, if the lepton energy spectra and the acceleration efficiencies of the msPSR populations are comparable. 
In this case the proportionality constant in Eq.~\ref{eq-IC-flux} can be fixed using the flux of \htfive , which then allows to calculate the expected VHE $\gamma$-ray flux for the other GCs. 

To estimate $N_\mathrm{msPSR}$ we used the stellar encounter rate $\Gamma_\mathrm{e} = \rho_0^{1.5} \times r_\mathrm{c}^2$ which is strongly correlated to the number of neutron star X-ray binaries, the progenitors of msPSRs \cite{2006ApJ...646L.143P}.
Here $\rho_0$ and $r_\mathrm{c}$ denote the central luminosity density and the core radius of the GC, respectively. 
To calculate $F_\mathrm{IC}$ we used the values for $\rho_0$, $d$ and $r_\mathrm{c}$ as given by \cite{1996AJ....112.1487H}. 
For $N_\mathrm{ph}$ we used four different target photon fields: 1) the optical light from the GC stars, 2) the Galactic infrared and 3) the Galactic optical background photon fields at the location of the GC (from GALPROP), and finally 4) the cosmic microwave background. 

To compare the expected IC flux with the upper limits we scaled the published VHE $\gamma$-ray flux from \htfive\ using Eq.~\ref{eq-IC-flux} by the ratios of each parameter for all invidiual GCs. 
Here we considered two extreme cases, namely that the leptons are either confined within the core region where the GC starlight is the dominant target photon field, or that the particles have propagated far away from the core where the Galactic background fields and the cosmic microwave background become the most important components. 
In the latter case we used the sum of the energy densities of the three photon fields as a proxy for $N_\mathrm{ph}$.
The ratio between the flux upper limits and the expected IC flux are given for both cases in the last two columns of Tab.~\ref{tab-analysis-results}. 
The uncertainties in this ratios originate from the limited accuracies of the measured input parameters. 

With the optical light of the GC member stars as the target photon field, the flux upper limits for point-like emission from the two GCs NGC\,6388 and NGC\,7078 lie about one order of magnitude below the expected IC flux.
Among all GCs studied here, given the uncertainties related to the predictions, only in these two cases the simple scaling model is challenged by our observational results. 
However, the fact that \htfive\ is extended and offset beyond the half-mass-radius of \terzan\ could point towards fast diffusion and long energy-loss timescales of the lepton population in \terzan . 
If this is indeed the case, then other target photon fields would be dominant over the GC starlight. 
Therefore, the results of the point-like analyses are better suited to be compared to the predictions with the GC starlight as target photon fields, whereas the results from the extended analyses are more relevant for the predictions based on Galactic background photons. 

To make a statement about the GC sample as a whole, we derived the expected IC flux from the total stack by calculating the weighted mean flux from all GCs. 
As weights we used the acceptance-corrected livetimes of the individual clusters. 
The upper limits for the stack are about two orders of magnitude below the expectation for the GC starlight and at a level of $\sim$50\% for the external target photon fields (see bottom section of Tab.~\ref{tab-analysis-results}). 
In the first case, even considering the uncertainties, this seems to be quite a strong constraint on the simple msPSR/IC scaling model. 
Therefore, either the IC flux from GCs does not scale as expected, or the VHE $\gamma$-ray source near \terzan\ is not related to msPSRs. 

There are many possibilities why \terzan\ may stand out among the general GC population. 
In the msPSR/IC model the confinement of leptons in the emission region might be stronger in \terzan\ compared to other GCs, which would lead to an accumulation of particles from earlier epochs that would therefore enhance the observed IC flux per msPSR. 
However, because \htfive\ is extended and offset from the core, the trapping mechanism is required to work on similarly large spatial scales. 
A high interstellar magnetic field strength in the vicinity of \terzan\ could provide such a trapping mechanism, but would also increase the synchrotron losses and therefore decrease the expected IC flux. 
These competing effects would need to be modeled in detail for each individual GC, and the results presented in this paper might yield interesting constraints. 

As already mentioned in sect.~\ref{sec-introduction} there are also models that are not related to the msPSR/IC scenario which suggest that the VHE $\gamma$-ray emission is due to hadronic processes in a short gamma-ray burst remnant \cite{2011A&A...533L...5D}. 
Due to the extremely low rates of these catastrophic events of only $\sim$10$^{-4}$\,yr$^{-1}$ \cite{2007PhR...442..166N} it could be easily explained why only one GC stands out among the whole population.

\end{document}